\documentstyle[aas2pp4]{article}

\received{28 February 1995}
\accepted{23 March 1995}

\slugcomment{To appear in Astro. \& Astrophys. Suppl. Series 114.}

\lefthead{Robert Braun}
\righthead{Resolved Atomic Super-clouds}

\begin{document}

\title{Resolved Atomic Super-clouds in Spiral Galaxies}

\author{Robert Braun}
\affil{Netherlands Foundation for Research in Astronomy, Postbus~2,
7990~AA Dwingeloo, The Netherlands}

\begin{abstract}
High quality data are presented of neutral hydrogen emission and
absorption in the fields of eleven of the nearest spiral galaxies.
Multi-configuration VLA observations have provided angular resolution of
6~arcsec (corresponding to about 100~pc at the average galaxy distance
of 3.5~Mpc) and velocity resolution of 6~km~s$^{-1}$, while accurately
recovering the total line flux detected previously with filled
apertures. Previous experience suggests that this physical resolution is
sufficient to at least marginally resolve the \ion{H}{1} super-cloud
population which delineates regions of active star formation. A high
brightness filamentary network of \ion{H}{1} super-clouds is seen in
each galaxy. Emission brightness temperatures in excess of 200~Kelvin
are sometimes detected at large radii, even in relatively face-on
systems. All galaxies display a systematic increase in the observed
brightness temperature of super-clouds with radius, followed by a
flattening and subsequent decline. In the few instances where
background continuum sources allow detection of \ion{H}{1} absorption,
the indicative spin temperatures are consistent with the super-cloud
brightness temperature seen in emission at similar radii. These data
suggest substantial opacity of the \ion{H}{1} in the super-cloud
network.
\end{abstract}


\section{Introduction}

It has become common practise to assume that the $\lambda$21~cm line of
neutral atomic hydrogen emission is in most cases an {\it optically
thin\ } tracer of the atomic gas mass in external galaxies. A number of
recent results are beginning to throw doubt on this conjecture. Careful
imaging studies of \ion{H}{1} absorption at low latitude in the Galaxy
reveal {\it highly} saturated absorption associated with the velocities
of spiral arms, with only lower limits to $\tau$ of 5 or 6 toward the
supernova remnant Cas~A (Bieging {\it et al.}, 1991) and the HII region
complex W43 (Liszt {\it et al.}, 1993). Such highly saturated
absorption is seen fairly uniformly over the several parsec spatial
extent of these sources. These high opacities can not be attributed to
velocity crowding along these lines-of-sight, since they are well
removed from the ``difficult'' longitudes of 0, 70 and 180$^\circ$,
where Galactic rotation and geometry conspire to give a minimal
gradient of line-of-sight velocity with Galactocentric distance. In
addition, the absorption profiles have a high degree of structure in
velocity, which is more consistent with an origin in discrete
components rather than a smoothly increasing opacity with path length
per velocity interval. Comparable \ion{H}{1} opacities are therefore
likely to be observed toward such spiral arm segments even if the
Galaxy were viewed face-on. Since only a few low latitude
lines-of-sight have received careful study, it is not yet possible to
estimate the prevalence of such regions. A program to significantly
extend the sample of well-imaged low latitude regions in \ion{H}{1}
absorption is now underway (Burton and Braun, 1995).

\subsection{Emission Brightness as an Opacity Tracer}

There is another, more accessible, tracer of the distribution of high
opacity neutral hydrogen. The emission brightness itself, T$_B$, is
tightly correlated with the absorption opacity, $\tau$, in the solar
neighborhood (Braun and Walterbos 1992, hereafter BW92). The simplest
interpretation of this correlation is that on physical scales of about
1~kpc, a single temperature of cool gas is prevalent. This is not to
say that only a single temperature {\it phase\ } of neutral gas is
present on these spatial scales. On the contrary, Galactic
lines-of-sight seem to exhibit as much as about 5~Kelvin of emission
without detectable associated absorption, implying the existence of a
widespread warm neutral phase with a kinetic temperature greater than
about 5000~K. However, it is striking that all lines-of-sight with more
than about 5~K of emission do have detectable associated absorption
suggesting that the warm neutral phase rarely attains a high column
density, at least in the solar neighborhood. Neither is it the case
that there is strict temperature equality of the cool gas over 1~kpc
scales. The solar neighborhood data are contained in an envelope
corresponding to a 50\% variation about a well-defined mean cool phase
temperature of about 110~K. This agrees with the peak brightness
temperature seen directly in emission along the local Galactic plane.
This general consistency is also borne out on a smaller scale. In a
study of some 50 cross-cuts spanning associated molecular and atomic
zones of optically identified Galactic dust lanes, Anderson {\it et
al.} (1991) see evidence that the high brightness, atomic hydrogen
envelopes of molecular clouds routinely have high \ion{H}{1} opacity
($\tau>1-2$) such that the {\it resolved\ } peak \ion{H}{1} brightness
temperature, T$_B$ is identical to the \ion{H}{1} kinetic temperature.
Furthermore, they observe (but do not comment on) higher \ion{H}{1}
kinetic temperatures in their outer Galaxy objects (R~$\sim$~11~kpc)
than in either the solar neighborhood or the inner Galaxy. A recent
analysis of the Galactic \ion{H}{1} emission properties based on the
Leiden/Dwingeloo survey (Braun, Burton and West 1996) has elucidated
the distribution of peak T$_B$ with radius in the Galaxy. A radial
increase in T$_B$ is observed out to about 10~kpc, with a subsequent
decline. Simulations of the Galactic data are most consistent with high
mean opacities of the gas over the radial interval over which the
temperature gradient is found. This same radial gradient of both the
peak T$_B$ together with the {\it identical\ } \ion{H}{1} kinetic
temperature is seen in M31 via HI absorption and associated emission
detected along multiple lines-of-sight (BW92) through the gaseous disk
of that system. Kinetic temperatures of about 70~K are typical at 5~kpc
radius in M31, while they increase to about 175~K at 15~kpc. The fact
that widespread regions are observed in M31 with a high emission
brightness in \ion{H}{1} which is comparable to the implied local
kinetic temperature suggests that regions of relatively high opacity in
\ion{H}{1} are similarly widespread.

\subsection{Resolving \ion{H}{1} Emission}

A necessary condition for observing the \ion{H}{1} kinetic temperature
of an opaque region directly as an emission brightness temperature is
that the region be resolved both spatially and in velocity. Since
structure in Galactic \ion{H}{1} is known to exist down to size scales
of less than a parsec, this may appear to be a rather elusive goal. The
physical resolution available to BW92 in the case of M31 was 33~pc
spatially and 5.2~km~s$^{-1}$. This is comparable to the physical
resolution in HI attained at the distance of the Galactic center with a
100~m filled aperture. Comparison of Galactic emission spectra at
various physical resolutions (eg. Fig.~2 of Dickey and Lockman 1990)
suggests that although detailed structures may certainly be
under-resolved with a larger beam, the peak brightness is not
substantially diluted even with 160~pc resolution (36~arcmin at
15~kpc). In a companion paper (Braun and Walterbos 1996) we consider in
detail the degree to which \ion{H}{1} emission structures in M31 have
been resolved with the resolution of BW92. It is shown that spatial
smoothing to 130~pc resolution gives a mean dilution of peak brightness
of only 8\% while velocity smoothing to 10~km~s$^{-1}$ gives rise to a
15\% mean dilution of peak brightness. The major emission complexes, or
atomic super-clouds, which give rise to high brightness \ion{H}{1}
emission regions apparently have typical dimensions larger than about
100~pc and line-widths which are only marginally resolved with
5~km~s$^{-1}$. This assessment is in very good agreement with that of
Elmegreen and Elmegreen (1987) who deduce a typical radius of 100~pc
for \ion{H}{1} super-clouds in the Galaxy and a typical velocity
dispersion of 5~km~s$^{-1}$.

The combination of spatial and velocity resolution necessary to resolve
the \ion{H}{1} super-clouds in galactic disks had never been applied to
external galaxies beyond Messier 31. The study of M31 (Braun, 1990 and
BW92) revealed higher brightnesses of \ion{H}{1} than ever before
detected in the Galaxy (up to 180~K) as well as the systematic increase
of the \ion{H}{1} kinetic temperature with radius noted above. What
other attributes and trends are waiting to be discovered in galaxies of
other morphological type? We set out to answer this question by
selecting a sample for resolved \ion{H}{1} imaging of the nearest
external galaxies beyond the local group. An effort was made to span a
large range of morphological type although the selection was limited
primarily by the need for proximity. This is because a sufficiently
high brightness sensitivity is necessary to actually detect the
\ion{H}{1} emission. For example, the A-configuration of the VLA could
provide angular resolution at $\lambda$21~cm of about 2~arcsec
(corresponding to 150~pc at 15~Mpc) but would achieve a brightness
sensitivity of only about 150~K per 5~km~s$^{-1}$ channel in an eight
hour integration. When limited to modest integration times of about
eight hours, the practical limiting angular resolution is the 6~arcsec
of the VLA B-configuration for which a brightness sensitivity of about
15~K is realized. This in turn imposes an upper limit to the distance
of about 5~Mpc to maintain a spatial resolution of better than 150~pc.

The observations and data reduction of our nearby galaxy sample are
described in \S~2 of this paper. This is followed by a brief
presentation of results in \S~3. The reader is referred to a companion
paper (Braun 1996, hereafter B96) for a more extensive analysis of the
data.

\section{Observations and Data Reduction}

Neutral hydrogen observations of the eleven program galaxies were
obtained with the VLA between March 1989 and November 1990. The B, C
and D configurations (with effective integration times of about 7, 0.5
and 0.4 hours) were utilized to image a region 0.5 degree in diameter
at 6 arcsec resolution for each of the eight Northern galaxies. In
addition, a small hexagonal mosaic in the D configuration (with 0.4
hours effective integration on each of 7 positions) was used to image a
1 degree diameter field at 65 arcsec resolution. Similar resolution and
sampling was obtained for the three southerly galaxies by observing in
the BnA, CnB and DnC configurations. Observing dates, field centers and
other particulars are summarized in Table 1. Namely, the galaxy name in
column (1), the B1950 pointing center in (2), the observation dates for
the three observed configurations in (3), (4) and (5), the central
velocity and number of frequency channels in (6) and (7). The assumed
inclination, position angle of receding line-of-nodes and major axis
radius at which the blue optical surface brightness is 25
mag~arcsec$^2$ (from de Vaucouleurs, de Vaucouleurs and Corwin, 1976,
hereafter the RC2) is given in columns (8), (9) and (10). The galaxy
type, approximate distance, total blue magnitude and luminosity are
given in columns (11) to (14). Standard calibration and imaging
techniques were used to produce a series of narrow-band images
separated by 5.16~km~s$^{-1}$ over a velocity range of 330~km~s$^{-1}$
(or 660~km~s$^{-1}$ when necessary) centered on the nominal
heliocentric systemic velocity of each galaxy. Since a uniform
frequency taper was applied in the correlator, the effective velocity
resolution was 6.2~km~s$^{-1}$. A continuum image, formed from the
average of line-free channels was subtracted from each image. A
simultaneous deconvolution based on the Maximum Entropy Method (as in
Cornwell 1988) was carried out on the seven pointings of the low
resolution mosaic to determine the smoothest model brightness
distribution at each velocity consistent with the data (corrected for
primary beam attenuation) and the measurement noise. The model
brightness distributions were then convolved with a two-dimensional
Gaussian fit to the synthesized beam to which were added the residuals
of the deconvolution. Since the zero level of each narrow-band image
remains unconstrained due to the absence of total power data, the mean
brightness in a region outside of the source was then subtracted from
each. The good response to extended structure (as large as about 30
arcmin) obtained in this low resolution mosaic was incorporated into
the series of high resolution images by replacing the inner Fourier
plane of each (as in Braun 1988). The high resolution images were not
subjected to deconvolution, since the good sampling, uniform weighting
and Gaussian tapering (at 25~k$\lambda$) already provided a very nearly
Gaussian instrumental response with a maximum near-in sidelobe level of
only a few percent. The only substantial short-coming of this database,
the so-called ``short spacing bowl'', was accurately removed by
including the low resolution mosaic data as discussed above. The image
brightness scale (in Kelvin) was defined on the basis of integrating
the actual instrumental response in elliptical annuli. This resulted in
a small correction (about 5\%) relative to the best fitting elliptical
Gaussian beam. The best fitting elliptical Gaussian beam parameters as
well as the beam integrals are listed in columns (2) and (3) of Table~2.

\subsection{Images of Peak Brightness}

Smoothed versions of the high resolution (6~arcsec) \ion{H}{1}
data-cubes were formed at resolutions of 9, 15, 25 and 65~arcsec. The
absolute flux density and surface brightness scales are estimated to be
accurate to better than 5\%. The rms sensitivity in a single frequency
channel is given in terms of flux density and surface brightness in
columns (4) and (5) of Table~2. The surface brightness sensitivity
after convolution to a circular Gaussian beam of 9, 15, 25 and
approximately 65~arcsec is given in columns (6) through (9) of the
table. Images were formed of the peak brightness observed along each
spatial pixel and the corresponding line-of-sight velocity. Images of
the peak brightness at the full spatial resolution are shown in panel
(a) of Figs.~1--10 for all program galaxies but NGC~4826. (Various
images of NGC~4826 have appeared previously in Braun et al. 1994.) The
noise properties of such peak brightness images are somewhat peculiar
and deserve some explanation. When no detectable signal is present in
the spectrum corresponding to any particular spatial pixel, a peak
value of about 3$\sigma$ is found from the 50 or so independent
velocity pixels which make up that spectrum. When a signal is present
at a level of greater than about 3$\sigma$ it is detected with an
uncertainty of 1$\sigma$. If the signal profile is broad relative to
the velocity resolution, a positive bias level will be introduced due
to the Gaussian noise statistics. It will be seen below that our
velocity resolution of 6.2~km~s$^{-1}$ is only sufficient to marginally
resolve the emission line profiles at high spatial resolution so that
the bias level of signals greater than about 4$\sigma$ is negligible.
Since only a single pointing position was observed for each galaxy at
high resolution, the peak brightness images show the radially
increasing 3$\sigma$ noise floor due to the primary beam correction
(corresponding approximately to a Gaussian of 30 arcmin FWHM). The
obvious implication is that the detection level is a function of
distance from the pointing direction. Since the typical 1$\sigma$ noise
at full spatial resolution is 18~K, the 3$\sigma$ noise floor climbs
from about 55~K at the pointing center to about 110~K at a radius of
1800~arcsec.

\subsection{Images of Integrated Emission}

Images of the integrated \ion{H}{1} emission were produced at various
resolutions by masking each cube with a positive 3$\sigma$ threshold
level of a lower resolution cube and then forming the sum. For example,
the 9~arcsec resolution integral images shown in panel (b) of
Figs.~1--10 were formed by imposing a mask based on a 3$\sigma$
threshold level in the 25~arcsec resolution cube. Only at the lowest
resolution of 65~arcsec, with masking by a 130~arcsec cube was the
total (un-masked) flux recovered. These low resolution integral images
are shown in panel (d) of Figs.~1--10. The use of a
3$\sigma$ threshold level insured that no more than the total flux was
obtained in the sum. In contrast, the use of a positive 2$\sigma$
threshold level, produces a positive bias in the blanked sum of more
than 20\% in excess of the total line flux.

\subsection{Images of Line-of-sight Velocity}

The line-of-sight velocity of the peak emission brightness at 65~arcsec
resolution is shown in panel (c) of Figs.~1--10. This method of
generating the projected velocity field is superior to the more commonly
used method of generating the first moment since it is insensitive to
the asymmetric profiles which are often observed at moderate spatial
resolutions for galaxies of non-zero inclinations. This same advantage
is gained over fits of a single Gaussian to the line profiles in
addition to offering a greater robustness in cases of low
signal-to-noise. With the high surface brightness sensitivity obtained
at this low spatial resolution it is possible to determine the velocity
field reliably out to the low column densities of the diffuse outer
disk.

\subsection{Emission Integrals}

Total detected line fluxes were determined by evaluating the integral
of the data cube at full spatial resolution within the region defined
by a contour drawn at a peak brightness of 4~K in the 65 arcsec
resolution peak brightness image. (Due to the method of data reduction
these integrals are necessarily identical at spatial resolutions of 9,
15 and 25~arcsec.) The integral was also evaluated within the
contiguous spatial region defined by the contour of zero integrated
intensity in the 65~arcsec resolution mosaic database. These two values
are listed in columns (10) and (11) of Table 2. In most cases 5 to 20\%
more line flux is detected with the second method. This is due to
both the presence of extremely diffuse gas (with less than 4~K
brightness) as well as the larger field of view of the mosaic database.
In the case of NGC~4826, the 4~K threshold method only allowed
detection of about 10\% of the total flux. Recent total power
measurements in the literature (as tabulated in Huchtmeier and Richter,
1989) generally lie between the values determined by the two methods
above.

\subsection{Search for \ion{H}{1} Absorption}

The continuum images of each field were searched for discrete sources
toward the \ion{H}{1} disks of the program galaxies. A two component
least squares fit was then made to small regions (about 3 beam-widths,
or 20~arcsec in diameter) in each continuum subtracted channel map to
solve for the amplitude of the elliptical Gaussian source
($1-e^{-\tau}$) and the amplitude of a constant interpolated background
level (T$_B$) as a function of velocity. This method of deriving
line-of-sight absorption and emission spectra was employed with success
by Braun and Walterbos (1992) in the case of M31. An important
difference between that study and the present one is the physical
resolution. The physical resolution available to BW92 for M31 was 35~pc
and the dimensions of the fitting region were about 100~pc. As
discussed in \S~1, the scale size of major emission regions in M31 is
about 100~pc, so that the fitting region was a reasonable match to this
scale-size. In the present study the physical resolution available for
our program galaxies was designed to be matched to this scale-size of
about 100~pc but our fitting region is necessarily about 300~pc in
extent. We therefore expect our derived spectra to have a higher
fluctuation level due to confusion effects resulting from variable
intensity levels within our fitting window. Inspection of the derived
absorption and emission spectra reveals that this is sometimes the
case. In these cases, the spatial baseline model was extended with
inclusion of first or second order polynomial terms. The rms
fluctuation level in the spectra in the presence of emission is still
somewhat increased over that expected statistically. In the many cases
of a slightly resolved continuum source this lost sensitivity is at
least partially compensated by the detection of the source over several
independent pixels.

\section{Results}

\subsection{General Emission Properties: The Super-Cloud Network }

Examination of the peak \ion{H}{1} brightness images in panel (a) of
Figs.~1--10 reveals how the distribution of atomic gas in our sample
galaxies becomes decomposed at 100~pc linear resolution into a high
brightness filamentary network of \ion{H}{1} super-clouds. The
continuity is best seen at low to moderate inclinations, while edge-on
systems like NGC~55 and 4244 appear entirely filled as seen in
projection. The super-cloud network corresponds globally to both the
grand-design, as well as the flocculent, spiral arms traced by massive
star formation and particularly by dust lanes in the various galaxies.
A worthwhile comparison is that with the (B-O) color image of NGC~5457
in Schweizer (1976, his Fig.~3g). An interesting trend seen in many
galaxies is for the lowest brightnesses to be observed at the smallest
radii. This is most obvious in systems like NGC~3031, 4736 and 5457
which have a high luminosity stellar disk, but is also seen in NGC~247,
2403 and 7793. Extreme values of less than about 50~K are found at kpc
radii in NGC~4736 for example. The opposite is also true, namely that
the highest brightnesses occur at large radii. Emission brightnesses of
150 to 200~K are observed at large radii in NGC~247, 3031, and 5457
with a clear systematic increase of brightness temperature with radius.
{\it In fact, all of the sample galaxies show the same systematic
behavior of the super-cloud brightness temperature with radius; an
initial increase out to some radius, a flattening and a subsequent
decline.}

\subsection{Absorption Properties}

The line-of-sight emission and absorption properties toward the 54
brightest continuum sources in the 11 observed fields are given in
Table 3. The galaxy field is indicated in column (1) of the table, the
B1950 position of each source in (2), the (deconvolved) major and minor
axis dimensions and major axis position angle in (3), the observed
values (without primary beam correction) of integrated and peak flux
density in (4) and (5), the primary beam attenuation factor at the
position of the source in (6), the rms error in the line-of-sight
optical depth in a single velocity channel based on the peak source
brightness in (7), the integral over the emission profile where the
brightness exceeds 5 K in (8), the velocity interval over which the
emission brightness exceeds 5 K in (9), the mean spin temperature or
3~$\sigma$ lower limit based on the ratio of emission to absorption in
(10) (3~$\sigma$ lower limits are calculated using the values of
integral emission and velocity width of columns 8, 9 together with the
$\sigma$ of column 7 divided by the square root of the number of
contributing velocity channels), the maximum emission brightness in
(11), the galactic disk radius intercepted by the line-of-sight in (12)
and a code in (13) which indicates whether the continuum source may be
intrinsic to the galaxy. Code R indicates a diffuse radio continuum
morphology, code H the presence of H$\alpha$ emission at the galaxy
redshift, code N indicates a nuclear source and a question mark
indicates some uncertainty in the galactic identification.

The small number of randomly distributed background sources brighter
than about 5~mJy/beam have only a very low probability of occurring
along a direction which intercepts the super-cloud network of HI
emission. The positions of the sources analyzed for the presence of
emission and absorption are overlaid on the images of peak HI
brightness and integrated emission in Figs.~1--10. Continuum sources
associated with recent massive star formation within the galaxies, such
as HII regions and supernova, while usually probing gas-rich
lines-of-sight, only rarely have a sufficient continuum brightness to
allow the detection of absorption. And even when they do, their
location within the galaxies confuses the interpretation, since only
the unknown fraction of the gas seen in emission which lies between us
and the continuum source can contribute to the absorption. Significant
absorption is only detected in a small number of cases. Spectra for the
lines-of-sight in NGC~55, 247, 2366, 2403 and 3031 are shown in
Fig.~11, while the spectrum for the NGC~4826 line-of-sight has appeared
previously in Braun et al. (1994). A high quality detection is made
only in NGC~247 toward what appears to be a true background source,
while 3 to 5 $\sigma$ detections are made toward HII region complexes
within NGC~55, 2403 and 4826. Only statistical detection of absorption
is seen toward a source intrinsic to NGC~2366 by summing the channels
with emission brightness greater than 5~K. Comparison of the source
locations with the peak brightnesses in panel (a) of Figs.~1--10
indicates that in all these cases but that of NGC~55, the continuum
sources are located within local minima of the \ion{H}{1} peak
brightness, and therefore do not probe the super-cloud opacity
directly. The bright nuclear source in NGC~3031 allows detection of
faint absorption (5\%) near a velocity of 0~km~s$^{-1}$ which is very
likely due to foreground gas in the Galaxy.

Comparison of column (10) of Table~3 with the images of peak brightness
in panel (a) of Figs.~1--10 suggests that the spin temperatures and
lower limits seen in absorption are consistent with the super-cloud
brightness temperatures seen in emission at similar radii in each
galaxy. Opacities toward the super-cloud network are therefore likely
to be substantial.

\subsection{Data Analysis}

As indicated in the Introduction, extensive analysis of the database
presented here can be found in a companion paper (B96). In that paper we
present many properties of the \ion{H}{1} super-cloud population as a
function of radius in the various galaxies, including the fractional
\ion{H}{1} line flux, the surface covering factor and measures of the
spatial and velocity compactness. We develop a detailed physical model
in terms of radial profiles of gas density and temperature which can
account for the observables quite successfully.

\acknowledgements

The generous allocation of VLA observing time by the NRAO to carry out
this project is gratefully acknowledged. It is a pleasure to
acknowledge the assistance of Sabrina di Grandi and Lucia Lazzari with
the extraction of absorption properties towards continuum sources in
the galaxy fields. The National Radio Astronomy Observatory is operated
by Associated Universities, Inc., under cooperative agreement with the
National Science Foundation.

\begin{deluxetable}{lccccccccccccc}
\tablecolumns{14}
\setlength{\tabcolsep}{2pt}
\scriptsize
\tablecaption{Log of Observations \label{tbl-1}}
\tablewidth{0pt}
\tablehead{
\colhead{Galaxy}      &\colhead{R.A.(1950), Dec.(1950)} &
\colhead{B or BnA}    &\colhead{C or CnB} &
\colhead{D or DnC}    &\colhead{V$_{Cen}$} &
\colhead{N$_{Ch}$}    &\colhead{Inc.} &
\colhead{PA}          &\colhead{R$_{25}$} &
\colhead{Type}        &\colhead{Dist.} &
\colhead{B$_T$}       &\colhead{L$_B$} \nl
\colhead{\ }          &\colhead{\ } &
\colhead{\ }          &\colhead{\ } &
\colhead{\ }          &\colhead{(km/s)} &
\colhead{\ }          &\colhead{($^\circ$)} &
\colhead{($^\circ$)}  &\colhead{(")} &
\colhead{\ }          &\colhead{(Mpc)} &
\colhead{(mag)}       &\colhead{(10$^9$L$_\odot$)} \nl
\colhead{(1)}          &\colhead{(2)} &
\colhead{(3)}          &\colhead{(4)} &
\colhead{(5)}          &\colhead{(6)} &
\colhead{(7)}          &\colhead{(8)} &
\colhead{(9)}          &\colhead{(10)} &
\colhead{(11)}          &\colhead{(12)} &
\colhead{(13)}          &\colhead{(14)}
}
\startdata
N55 &00 12 24.00 $-$39 28 00.0 &N/A &20-05-89 &20-10-89 &+140 &64 &80
&110 &972 &Sc &2.0 &8.22 &3.2 \nl
N247 &00 44 40.00 $-$21 02 24.0
&04/05-03-89 &20-05-89 &20-10-89 &+160 &64 &75 &170 &600 &Sc &2.5 &9.51
&1.5 \nl
N2366 &07 23 37.00 +69 15 05.0 &19/28-03-89 &29-11-90
&01-12-89 &+100 &64 &58 &39 &228 &SBm &3.3 &11.46 &0.44 \nl
N2403 &07
32 01.20 +65 42 57.0 &19-03-89 &29-11-90 &01-12-89 &+130 &64 &62 &125
&534 &Sc &3.3 &8.89 &4.7 \nl
N3031 &09 51 27.60 +69 18 13.0 &16-03-89
&29-11-90 &01-12-89 &$-$45 &128 &60 &330 &774 &Sb &3.3 &7.86 &12. \nl
N4236 &12 14 21.80 +69 44 36.0 &28-03-89 &29-11-90 &01-12-89 &0 &64 &73
&161 &558 &SBd &3.3 &10.06 &1.6 \nl
N4244 &12 14 59.90 +38 05 06.0
&23-03/02-05-89 &29-11-90 &30-11-89 &+245 &64 &80: &233 &486 &Scd &4
&10.60 &1.4 \nl
N4736 &12 48 32.40 +41 23 28.0 &30-03-89 &29-11-90
&30-11-89 &+305 &64 &33 &305 &330 &Sab &3.8 &8.92 &6.1 \nl
N4826 &12 54
16.90 +21 57 18.0 &22-03-89 &29-11-90 &30-11-89 &+415 &128 &66 &300
&282 &Sab &3.8 &9.37 &4.0 \nl
N5457 &14 01 26.60 +54 35 25.0 &21-03-89
&29-11-90 &01-12-89 &+230 &64 &27 &40 &810 &Sc &6.5 &8.18 &35. \nl
N7793 &23 55 16.00 $-$32 52 06.0 &04/05-03-89 &20-05-89 &20-10-89 &+230
&64 &48 &289 &276 &Sd &3.4 &9.65 &2.5 \nl
\enddata
\end{deluxetable}

\begin{deluxetable}{lcccccccccc}
\tablecolumns{11}
\setlength{\tabcolsep}{6pt}
\scriptsize
\tablecaption{Data Attributes and Results \label{tbl-2}}
\tablewidth{0pt}
\tablehead{
\colhead{Galaxy} &\colhead{a b p} &\colhead{$\Omega_B$} &
\colhead{$\Delta$S$_F$}    &\colhead{$\Delta$T$_F$} &
\colhead{$\Delta$T$_9$}    &\colhead{$\Delta$T$_{15}$} &
\colhead{$\Delta$T$_{25}$} &\colhead{$\Delta$T$_{65}$} &
\colhead{$\int F_4 dV$}    &\colhead{$\int F_0 dV$} \nl
\colhead{\ }               &\colhead{(") (") ($^\circ$)} &
\colhead{(a.s.$^2$)}       &\colhead{(mJy/bm)} &
\colhead{(K)}              &\colhead{(K)} &
\colhead{(K)}              &\colhead{(K)} &
\colhead{(K)}              &
\multicolumn{2}{c}{(Jy-km~s$^{-1}$)} \nl
\colhead{(1)}          &\colhead{(2)} &
\colhead{(3)}          &\colhead{(4)} &
\colhead{(5)}          &\colhead{(6)} &
\colhead{(7)}          &\colhead{(8)} &
\colhead{(9)}          &\colhead{(10)} &
\colhead{(11)}
}
\startdata
N55 &16.0 12.2 $-$15 &220 &3.5 &10.9 &... &... &3.98 &0.93 &1525 &1525
\nl
N247 &6.58 6.01 $-$10 &61.6 &1.9 &21.1 &12.7 &6.47 &3.78 &0.76 &765 &860
\nl
N2366 &5.67 5.66 $-$40 &33.8 &1.3 &26.4 &11.2 &4.58 &1.94 &0.73 &235 &250
\nl
N2403 &6.25 6.12 +66 &48.0 &1.2 &17.2 &9.73 &4.85 &2.52 &0.68 &1320 &1440
\nl
N3031 &6.13 5.84 $-$1 &45.6 &1.1 &16.6 &8.98 &4.31 &2.52 &0.70 &1455 &1865
\nl
N4236 &5.93 5.79 $-$38 &49.2 &1.2 &16.8 &8.98 &4.85 &2.52 &0.63 &550 &610
\nl
N4244 &6.84 5.95 $-$82 &55.6 &1.4 &17.4 &10.5 &5.12 &2.62 &0.57 &410 &435
\nl
N4736 &6.55 5.87 +85 &48.8 &1.3 &18.3 &9.73 &4.58 &2.23 &0.58 &43 &70
\nl
N4826 &6.49 6.10 $-$67 &46.4 &1.3 &19.2 &9.73 &4.58 &2.23 &0.58 &... &47
\nl
N5457 &6.04 5.86 +69 &44.8 &1.1 &16.8 &8.23 &4.31 &2.43 &0.62 &1495 &1880
\nl
N7793 &6.98 6.06 $-$5 &53.2 &2.0 &26.0 &17.2 &8.09 &3.69 &0.82 &210 &270
\nl
\enddata
\end{deluxetable}

\clearpage
\begin{deluxetable}{lcccccccccccc}
\tablecolumns{13}
\setlength{\tabcolsep}{2pt}
\scriptsize
\tablecaption{HI absorption data \label{tbl-3}}
\tablewidth{0pt}
\tablehead{
\colhead{Galaxy}        &\colhead{R.A.(1950), Dec.(1950)} &
\colhead{a b p}         &\colhead{S$_{tot}$} &
\colhead{S$_{pk}$}      &\colhead{PBa} &
\colhead{$\sigma_\tau$} &\colhead{$\int T_BdV$} &
\colhead{$\Delta$V}     &\colhead{$<T_{sp}>$} &
\colhead{T$_{B max}$}   &\colhead{R$_{gal}$} &\colhead{Cl} \nl
\colhead{\ }            &\colhead{\ } &
\colhead{(") (") ($^\circ$)} &\colhead{(mJy)} &
\colhead{(mJy/bm)}      &\colhead{\ } &
\colhead{\ }            &\colhead{K km s$^{-1}$} &
\colhead{km s$^{-1}$}   &\colhead{(K)} &
\colhead{(K)}           &\colhead{(arcsec)} &\colhead{\ } \nl
\colhead{(1)}          &\colhead{(2)} &
\colhead{(3)}          &\colhead{(4)} &
\colhead{(5)}          &\colhead{(6)} &
\colhead{(7)}          &\colhead{(8)} &
\colhead{(9)}          &\colhead{(10)} &
\colhead{(11)}          &\colhead{(12)} &
\colhead{(13)}
}
\startdata
N55 &00 12 26.98 $-$39 29 00.8 &26.3 9.0 150 &19.93 &8.22 &1.00 &0.43 &4836
&92.9 &140$\pm$40 &140 &49: &R \nl
\ &00 12 29.81 $-$39 29 14.6 &38.3 27.4 156
&52.09 &8.09 &1.00 &0.43 &4754 &87.7 &$>$175 &100 &52: &R \nl
\ &00 12 43.79
$-$39 27 26.3 &0.0 0.0 0 &25.57 &25.78 &1.03 &0.14 &544 &77.4 &$>$65 &20.3
&308: &... \nl
N247 &00 44 21.01 $-$20 58 53.7 &6.5 5.9 81 &7.47 &3.77 &1.07
&0.50 &12.9 &5.2 &$>$2 &2.7 &931 &... \nl
\ &00 44 44.88 $-$21 07 36.2 &6.0
2.9 154 &47.4 &31.4 &1.06 &0.06 &1313 &87.7 &63$\pm$15 &38.9 &349 &...
\nl
\ &00 44 53.89 $-$21 17 45.0 &5.9 5.2 148 &39.8 &22.4 &2.01 &0.08 &21.3 &10.3
&$>$12 &2.7 &980 &... \nl
\ &00 44 56.41 $-$21 05 50.1 &0.0 0.0 0 &5.62 &5.07
&1.06 &0.37 &84.1 &25.8 &$>$7 &4.7 &782 &... \nl
N2366 &07 23 25.17 +69 17
29.0 &4.5 4.4 101 &8.70 &5.38 &1.00 &0.24 &1748 &77.4 &145$\pm$60 &65 &76 &H
\nl
\ &07 25 06.93 +69 12 03.9 &1.7 1.5 133 &9.48 &8.79 &1.36 &0.15 &0 &0 &...
&1.8 &1822: &... \nl
N2403 &07 29 29.48 +65 34 33.7 &6.6 2.6 52 &19.26 &12.12
&2.45 &0.10 &13.3 &5.2 &$>$9 &2.6 &2091 &... \nl
\ &07 30 19.23 +65 59 44.7
&9.1 2.1 54 &17.20 &9.19  &3.06 &0.13 &0 &0 &... &1.5 &1499 &... \nl
\ &07 30
24.47 +65 46 23.9 &2.5 2.2 59 &10.53 &9.22 &1.35 &0.13 &419 &56.8 &$>$63 &30
&739 &... \nl
\ &07 32 18.37 +65 43 21.3 &14.5 5.1 81 &10.27 &3.12 &1.00 &0.38
&1198 &98.0 &53$\pm$15 &45.2 &171 &H \nl
\ &07 32 41.72 +65 36 24.2 &2.6 2.4
167 &9.34 &8.02 &1.16 &0.15 &409 &41.3 &$>$62 &25 &554 &... \nl
\ &07 33 02.14
+65 47 14.0 &2.2 1.0 77 &11.13 &10.36 &1.15 &0.12 &223 &36.1 &$>$45 &25 &910
&... \nl
\ &07 33 09.08 +65 55 47.5 &5.0 2.5 71 &7.34 &5.29 &1.84 &0.23 &0 &0
&... &2.0 &1852 &... \nl
\ &07 33 09.89 +65 55 52.3 &7.4 2.6 116 &8.92 &5.29
&1.85 &0.23 &0 &0 &... &1.8 &1866 &... \nl
\ &07 33 56.27 +65 44 34.5 &3.3 1.8
63 &29.49 &25.05 &1.50 &0.05 &109 &25.8 &$>$63 &6 &1146 &... \nl
N3031 &09 49
02.49 +69 17 59.2 &7.0 6.4 11 &33.77 &14.99 &1.60 &0.07 &89.7 &25.8 &$>$37 &4
&1394 &... \nl
\ &09 49 04.03 +69 18 11.3 &13.5 6.4 43 &31.39 &8.63 &1.58
&0.13 &87.8 &25.8 &$>$20 &4 &1371  &... \nl
\ &09 50 05.52 +69 13 02.5 &3.5
2.3 169 &5.39 &4.36 &1.23 &0.25 &44.5 &15.5 &$>$7 &3 &1067 &... \nl
\ &09 50
36.38 +69 31 57.2 &5.5 4.8 174 &10.00 &5.76 &1.82 &0.19 &178 &25.8 &$>$27 &10
&1100 &... \nl
\ &09 51 12.96 +69 24 04.4 &0.0 0.0 0 &3.13 &3.04 &1.09 &0.36
&469 &56.8 &$>$25 &22 &406 &H? \nl
\ &09 51 27.32 +69 18 08.3 &2.3 2.0 141
&90.28 &79.87 &1.00 &0.014 &17.7 &5.2 &$>$80 &2.5 &0.0 &N \nl
\ &09 52 46.82
+69 11 12.2 &0.0 0.0 0 &4.25 &3.79 &1.31 &0.29 &213 &41.3 &$>$17 &10 &654 &...
\nl
\ &09 53 42.81 +69 24 48.7 &5.8 3.6 154 &23.80 &14.71 &1.70 &0.07 &81.7
&25.8 &$>$34 &3 &1634 &... \nl
\ &09 53 43.57 +69 25 05.7 &6.1 3.6 32 &16.97
&10.15 &1.72 &0.11 &64.2 &20.6 &$>$19 &3 &1658 &... \nl
N4236 &12 13 03.90 +69
48 14.6 &6.6 3.3 46 &6.94 &4.01 &1.15 &0.30 &0 &0 &... &2 &1113 &... \nl
\ &12
13 15.06 +69 50 22.3 &3.1 2.7 133 &8.21 &6.60 &1.18 &0.18 &45.1 &15.5 &$>$9 &6
&852 &... \nl
\ &12 14 13.97 +69 44 40.2 &0.0 0.0 0 &5.16 &4.70 &1.00 &0.26
&825 &61.9 &$>$59 &33 &128 &H? \nl
\ &12 14 14.07 +69 50 16.4 &0.0 0.0 0 &7.22
&6.71 &1.07 &0.18 &485 &51.6 &$>$55 &28 &418 &? \nl
\ &12 14 18.23 +69 45 41.1
&4.5 3.9 153 &12.0 &8.00 &1.00 &0.15 &863 &72.2 &$>$99 &47 &69 &H \nl
N4244
&12 13 52.65 +37 49 20.9 &9.8 3.4 91 &10.9 &5.39 &3.40 &0.26 &0 &0 &... &2
&4112: &... \nl
\ &12 14 28.72 +38 11 26.3 &7.7 7.1 5 &14.0 &5.95 &1.22 &0.24
&0 &0 &... &2 &7529: &... \nl
\ &12 15 40.47 +38 16 34.2 &6.0 3.8 146 &10.1
&6.31 &1.74 &0.22 &0 &0 &... &1.5 &3845: &... \nl
N4736 &12 48 31.93 +41 23
31.5 &6.7 5.1 176 &18.05 &9.36 &1.00 &0.14 &137 &46.4 &$>$21 &4 &0.0 &N
\nl
\ &12 48 35.59 +41 23 24.3 &11.1 5.0 178 &7.69 &2.87 &1.00 &0.45 &430 &87.7
&$>$15 &20 &38 &H \nl
N4826 &12 53 43.86 +21 59 48.9 &7.9 7.7 6 &5.68 &2.25
&1.17 &0.58 &0 &0 &... &2 &519 &... \nl
\ &12 53 51.67 +21 58 38.6 &0.0 0.0 0
&21.0 &13.6 &1.08 &0.10  &0 &0 &... &2 &415 &... \nl
\ &12 54 05.77 +21 44
19.7 &0.0 0.0 0 &10.0 &9.93 &1.65 &0.13 &0 &0 &... &1.5 &1843 &... \nl
\ &12
54 16.28 +21 57 12.7 &17.0 12.7 110 &53.9 &8.31 &1.00 &0.16 &987 &129
&60$\pm$10 &14 &2 &H \nl
\ &12 54 23.23 +22 09 42.4 &0.0 0.0 0 &7.15 &6.81
&1.55 &0.19 &0 &0 &... &1.5 &1740 &... \nl
\ &12 54 32.26 +22 08 28.1 &2.3 2.0
15 &9.58 &8.56 &1.47 &0.15 &0 &0 &... &1 &1718 &... \nl
\ &12 54 32.50 +22 08
50.1 &4.6 2.4 18 &6.20 &4.62 &1.52 &0.28 &0 &0 &... &1.5 &1770 &... \nl
\ &12
54 52.59 +21 45 19.3 &0.0 0.0 0 &7.22 &7.18 &1.84 &0.18 &0 &0 &... &1.5 &1197
&... \nl
N5457 &14 00 01.08 +54 43 17.1 &6.8 4.3 76 &33.3 &17.8 &1.85 &0.06 &0
&0 &... &2 &985 &... \nl
\ &14 01 14.69 +54 28 49.5 &12.2 9.3 17 &6.40 &1.51
&1.12 &0.71 &735 &51.6 &$>$20 &38 &418 &H \nl
\ &14 01 31.59 +54 36 21.9 &7.2
7.0 113 &10.1 &4.18 &1.00 &0.26 &191 &36.1 &$>$18 &20 &72 &... \nl
\ &14 01
32.65 +54 36 15.4 &6.4 5.6 168 &7.80 &3.87 &1.00 &0.28 &197 &36.1 &$>$17 &20
&73 &... \nl
\ &14 01 39.13 +54 44 47.7 &0.0 0.0 0 &4.36 &4.31 &1.28 &0.26
&191 &31.0 &$>$19 &30 &590 &... \nl
\ &14 01 55.35 +54 33 26.5 &10.6 6.0 44
&16.8 &5.84 &1.04 &0.19 &872 &67.1 &$>$82 &72 &308 &H \nl
\ &14 02 09.48 +54
23 05.8 &3.8 2.5 71 &5.83 &4.53 &2.32 &0.24 &0 &0 &... &2 &929 &... \nl
\ &14
02 15.94 +54 27 04.9 &5.5 4.2 146 &8.92 &5.37 &1.40 &0.20 &60.0 &20.6 &$>$10
&4 &738 &... \nl
\ &14 02 43.39 +54 38 09.4 &9.5 7.0 148 &8.94 &3.08 &1.44
&0.36 &814 &56.8 &$>$44 &68 &717 &H \nl
N7793 &23 54 56.02 $-$32 54 09.7 &5.7
4.7 0 &6.67 &4.09 &1.04 &0.49 &303 &61.9 &$>$12 &7 &866 &? \nl
\enddata
\end{deluxetable}

\clearpage

\figcaption{Images of NGC~55. {\bf (a).} Peak brightness of neutral
hydrogen emission as observed with full spatial and velocity resolution
(about 100~pc and 6~km~s$^{-1}$). {\bf (b).} Integrated neutral
hydrogen emission at about 150~pc spatial (9 arcsec, except 15 arcsec
for NGC~55) resolution. Masking for the integral was carried out at the
3$\sigma$ contour at about 400~pc (25 arcsec) resolution. {\bf (c).}
The line-of-sight velocity of the peak brightness at 65~arcsec
resolution. Iso-velocity contours are drawn at intervals of
15~km~s$^{-1}$.{\bf (d).} Integrated neutral hydrogen emission at
65~arcsec resolution. The integrals have been converted to apparent
column density under the usual (but incorrect) assumption of negligible
optical depth. Contours are drawn at 1, 2, 4, 8, 16, 32 and 64 in units
of 10$^{20}$cm$^{-2}$. As discussed in the B96, actual column densities
are likely to be locally enhanced on spiral arm segments by factors of
several. The linear grey scale has lower and upper limits as indicated.
The filled cross marks the position of the pointing center. The
un-filled crosses mark the position of continuum sources against which
the data were analyzed for the presence of absorption. Note the trend
for an increasing peak \ion{H}{1} brightness with radius in the
galaxies NGC~247, 3031 and 5457 as well as the extremely low peak
brightnesses observed at small radii in NGC~4736. \label{fig1}}

\figcaption{Images of NGC~247 with panels as in Fig.~1. \label{fig2}}

\figcaption{Images of NGC~2366 with panels as in Fig.~1. \label{fig3}}

\figcaption{Images of NGC~2403 with panels as in Fig.~1. \label{fig4}}

\figcaption{Images of NGC~3031 with panels as in Fig.~1. \label{fig5}}

\figcaption{Images of NGC~4236 with panels as in Fig.~1. \label{fig6}}

\figcaption{Images of NGC~4244 with panels as in Fig.~1. \label{fig7}}

\figcaption{Images of NGC~4736 with panels as in Fig.~1. \label{fig8}}

\figcaption{Images of NGC~5457 with panels as in Fig.~1. \label{fig9}}

\figcaption{Images of NGC~7793 with panels as in Fig.~1. \label{fig10}}

\figcaption{Line-of-sight \ion{H}{1} absorption (top) and inferred
emission (middle) toward continuum sources in Table ~3 with (tentative)
detections of absorption. The method of spectrum extraction is
described in the text. A low spatial resolution (65 arcsec) emission
spectrum is included (bottom panel) for reference. \label{fig11}}

\end{document}